\documentclass[11pt]{article}
\newcommand{\Comment}[1]{{}}
\usepackage{amsfonts,amsthm,amsmath,amssymb,slashed}
\usepackage[textwidth = 430 pt, textheight = 630 pt]{geometry}
\usepackage{color}

\Comment{\usepackage{color}
\definecolor{MyDarkBlue}{rgb}{0.15,0.15,0.45}
\usepackage[linktocpage=true]{hyperref}
\hypersetup{
colorlinks=true,
citecolor=MyDarkBlue,
linkcolor=MyDarkBlue,
urlcolor=MyDarkBlue,
pdfauthor={Horatiu Nastase},
pdftitle={Quantum gravity and the holographic dark energy},
pdfsubject={hep-th}
}

\usepackage[numbers,sort&compress]{natbib}
\usepackage{hypernat}}
\usepackage{graphicx}
\usepackage{cite}

\newcommand\ignore[1]{}
\def\one{{\,\hbox{1\kern-.8mm l}}}

\def\a{\alpha}

\newcommand{\Cset}{{\,\,{{{^{_{\pmb{\mid}}}}\kern-.45em{\mathrm C}}}}}

\newcommand{\be}{\begin{equation}}
\newcommand{\bea}{\begin{eqnarray}}

\newcommand{\ee}{\end{equation}}
\newcommand{\eea}{\end{eqnarray}}

\providecommand{\lsim}{\lesssim}

\begin{document}

\renewcommand{\thefootnote}{\fnsymbol{footnote}}

\makeatletter
\@addtoreset{equation}{section}
\makeatother
\renewcommand{\theequation}{\thesection.\arabic{equation}}

\rightline{}
\rightline{}


\vspace{10pt}


\begin{center}
{\LARGE \bf{\sc Quantum gravity and the holographic dark energy cosmology}}
\end{center} 
 \vspace{1truecm}
\thispagestyle{empty} \centerline{
{\large \bf {\sc Horatiu Nastase}}\footnote{E-mail address: \Comment{\href{mailto:nastase@ift.unesp.br}}{\tt nastase@ift.unesp.br}}
                                                  }

\vspace{.5cm}


\centerline{{\it 
Instituto de F\'{i}sica Te\'{o}rica, UNESP-Universidade Estadual Paulista}} \centerline{{\it 
R. Dr. Bento T. Ferraz 271, Bl. II, Sao Paulo 01140-070, SP, Brazil}}

\vspace{1truecm}

\thispagestyle{empty}

\centerline{\sc Abstract}

\vspace{.4truecm}

\begin{center}
\begin{minipage}[c]{380pt}
{\noindent The holographic dark energy model is obtained from a cosmological constant generated by generic quantum gravity effects giving a 
minimum length. By contrast, the usual bound for the energy density to be limited by the formation of a black hole simply gives the Friedmann equation. 
The scale of the current cosmological constant relative to the inflationary scale is an arbitrary parameter characterizing initial conditions, 
which however can be fixed by introducing a physical principle during inflation, as a function of the number of e-folds and the inflationary scale.
}
\end{minipage}
\end{center}

\vspace{.5cm}

\setcounter{page}{0}
\setcounter{tocdepth}{2}

\newpage

\renewcommand{\thefootnote}{\arabic{footnote}}
\setcounter{footnote}{0}

\linespread{1.1}
\parskip 4pt



\section{Introduction}
\ \ \ \ \
The cosmological constant problem is one of the most important problems in theoretical physics (see, e.g. \cite{Padmanabhan:2002ji} for a review). 
On the one hand, we know that in quantum field 
theory there should be a vacuum energy coming from bubble Feynman diagrams, or equivalently  by summing over zero-point energies 
of the mode oscillators, $\sum_n\hbar \omega_n/2$. The value of the resulting cosmological constant energy density $\rho_\Lambda$
should be of the order of $M^4$, where $M$ is the maximum cut-off of the theory. In a generic theory including quantum gravity, this would be at the 
Planck scale, $M_{\rm Pl}^4$. Since a fermionic degree of freedom gives minus the above result for a bosonic one, supersymmetry (in the absence of 
gravity) would imply a zero cosmological constant, but a theory with broken supersymmetry like the real world would give $\rho_\Lambda\sim
M_{SSB}^4$, where $M_{SSB}$ is the scale of supersymmetry breaking, and moreover in supergravity even a supersymmetric vacuum would 
have a (negative) cosmological constant.
On the other hand, cosmologically it was experimentally found that we have a "dark energy" component in the make-up of the Universe
(see the reviews \cite{Frieman:2008sn,Caldwell:2009ix,Li:2011sd}), which is 
described to a large degree of accuracy by a cosmological constant composing more than 2/3 of the total energy. 
According to the recent Planck data, we have a flat Universe ($\Omega_{\rm total}\simeq 1$), with 
$\Omega_\Lambda=0.6911\pm 0.0062$ and an equation of state $w_\Lambda=-1.006\pm 0.045$ (assuming $w$ is constant) \cite{Ade:2015xua}.
This possible cosmological constant however would be about 122 orders of magnitude smaller than its natural Planck scale value, 
$\rho_\Lambda\sim 10^{-122}M_{\rm Pl}^4$, which means either an incredible fine-tuning, which most physicists would dismiss, or the effect 
of some unknown mechanism. Note that since the Friedmann equation can be written in the form
\be
\rho_{\rm total}=3M_{\rm Pl}^2 H^2\;,\label{friedmann}
\ee
where $M_{\rm Pl}$ is the reduced Planck mass $M_{\rm Pl}=(8\pi G)^{-1/2}$,
the cosmological constant energy density is of the order $\rho_\Lambda\sim M^2_{\rm Pl}/(H^{-1})^2$.

A possible way out of this problem was suggested by the observation in \cite{Cohen:1998zx} that imposing that the total energy of vacuum fluctuations
in a size $L$ is smaller than the one required to create black hole of radius $L$ severely restricts the energy on cosmological scales, giving 
$\rho_\Lambda L^3\lsim L M_{\rm Pl}^2$. We will come back to discuss in more detail this bound later. Then, in \cite{Hsu:2004ri} an attempt was made 
to derive a consistent cosmology assuming that $L$ is given by the horizon size $H^{-1}$, which however contradicts experiment. 
A consistent cosmology was developed in \cite{Li:2004rb}, 
assuming that the correct scale $L$ to be considered is the future event horizon $R_h=a\int_t^\infty dt/a$ (after discarding as a possibility the 
particle horizon $R_H=a\int_0^t dt/a$), which was named "holographic dark energy" cosmology (for recent developments in holographic dark energy 
cosmology, see \cite{Li:2009bn,Li:2009zs,Wang:2012uf}. The model was later refined to introduce an 
interaction with dark matter in \cite{Pavon:2005yx,Wang:2005jx}.
In this letter I will argue that the "holographic dark 
energy" type of cosmological constant can naturally arise from a generic quantum gravity theory, assuming only the existence of a minimum length. 
I will then also expand on the model in  \cite{Li:2004rb}, presenting more details about the time evolution in the context of an inflationary cosmology, 
and show that the addition of a simple physical principle can fix the ratio of the cosmological constant today to the one during inflation in terms of 
the number of e-folds of inflation.

\section{Cosmological constant}

In a theory without gravity, the vacuum energy $\sum_n \hbar \omega_n/2$ presents an observable effect only in its modification due to the variation 
of the geometry of space, like in the well-known Casimir effect (see, e.g. \cite{Plunien:1986ca,Bordag:2001qi} for a review). Boundary conditions 
on an electromagnetic field, given by the presence of a conductor of a given geometry (e.g., infinite parallel plates, or sphere, etc.), can 
be varied, and the difference in the vacuum energy for different boundary conditions is an experimentally measurable effect. So if effect, what we 
measure is $\sum_n(\hbar \omega_n-\hbar\omega_{n,0})/2$, with $\hbar \omega_n/2$ the vacuum energy in the presence of the boundary conditions
and $\hbar \omega_{n,0}/2$ the vacuum energy for free space. This is in fact part of a more general technique to determine quantum corrections 
to the energy of a state. One considers the (regularized and renormalized) difference in vacuum energy in the presence of the state, for instance a 
soliton, with respect to the vacuum. This method for solitons is described for instance in the textbook \cite{Rajaraman:1982is}. In the case of two-dimensional
solitons, it was started by the classic work of \cite{Dashen:1974ci,Dashen:1974cj,Dashen:1975xh}, but since then, the calculation had a long history, because 
of the fact that there are many subtleties related to regularization and renormalization (some of the issues were fixed in \cite{Nastase:1998sy}, 
but others still occupy current research). The energy of the vacuum however (the potentially infinite additive constant subtracted above)
is considered in both of these (related) cases (Casimir effect and 
quantum corrections to the mass of solitons)  to not be unobservable, since it is an absolute constant that does affect physical processes. 

As soon as we consider the coupling to (classical) gravity however, the constant would gravitate and have observable effects, so we have to wonder why is it 
that we observe the differences in vacuum energy (Casimir effect and quantum corrections to the mass of solitons), but not the additive constant. 
We might however take it as a {\em physical principle} that {\em we cannot observe any absolute additive constant, but only changes due to the geometry of 
space or of the field configuration}. But what if we consider {\em quantum gravity} to be part of the theory as well (not simply classical general relativity)?
That will also change the "geometry of space" or the "field configuration" from the one with a fixed gravitational background. 

It is known for a long time now that generically, in the presence of any type of quantum gravity theory, we can describe some of its effects on spacetime
through a modification of the quantum uncertainty relation $\Delta x \Delta p\geq \hbar /2$ to something that admits a minimum length, usually 
expressed as 
\be
\Delta x\Delta p\geq \frac{1}{2}\left(\hbar +\frac{\a}{M_{\rm Pl}^2} \Delta p^2\right)\;,
\ee
called generalized uncertainty relation,
though other forms are also possible. Note that we have put explicitly $\hbar $ to emphase the quantum nature of the relation.
I will not review the long history of these ideas, since it is very well described in the review \cite{Hossenfelder:2012jw},
to which I refer the interested reader. A natural value for $\a$ would be 1 in a pure quantum gravity theory, since if we replace the quantum uncertainty
$\Delta x$ with a length scale $x$ and the uncertainty $\Delta p$ with a matching momentum scale $p$, we would obtain
\be
x\sim \frac{1}{2}\left(\frac{\hbar}{p}+\frac{\a}{M_{\rm Pl}^2}p\right)\;,
\ee
which has a minimum for $p_m=M_{\rm Pl}\sqrt{\hbar/\a}$, giving a minimal length
\be
x_{\rm min}=\frac{\sqrt{\hbar \a}}{M_{\rm Pl}}.
\ee
Thus in a pure quantum gravity theory, $\a=1$ would be the natural choice.
In string theory, one finds that strings at trans-Planckian energies tend to spread out over a large distance, an effect that can be 
understood in terms of of a minimum length and a modified uncertainty relation as above. One  of the early papers describing this is 
\cite{Amati:1988tn} (after the original papers on trans-Planckian scattering in string theory by Gross and Mende \cite{Gross:1987kza,Gross:1987ar}), 
and we can see that then the natural scale appearing in the uncertainty relation is the string scale, $\a/M_{\rm Pl}^2=\a'\equiv 1/M_s^2$. 
Since the string scale is usually smaller than the Planck scale by factors depending on the couplings (and maybe the volume of compactification, 
depending on models), in string theory we would naturally  have $\a<1$. Note that it has been argued by Yoneya that the uncertainty relation in string theory 
is modified differently than in the above \cite{Yoneya:2000bt} due to some non-perturbative effects, but we will nevertheless keep using it in this form.

In light of this generalized uncertainty relation, we model the effect of the change due to quantum gravity on the "geometry of space" or "field 
configuration" by replacing regular length scales 
\be
L\equiv 2x_0 =\frac{\hbar}{p}
\ee
with length scales admiting a minimum, 
\be
\tilde L\equiv 2x =\frac{\hbar}{p}+\frac{\a }{M_{\rm Pl}^2}p=L+\frac{\a\hbar}{M_{\rm Pl}^2L}.
\ee
We will revert to putting $\hbar=1$ from now on.

We now want to apply this to the problem of the cosmological constant. According to our physical principle, we can only observe changes due to 
the modification of the "geometry of space" or "field configuration", which in the case of quantum gravity we have argued that correspond to replacing 
length scales $L$ with length  scales $\tilde L$. Consider then the vacuum energy density due to quantum gravity
for a cubic system of size $L$, with momenta $k_{n_i}=n_i/L$ in the 3 spatial directions, $i=1,2,3$. The graviton zero point energies (with 2 physical 
polarizations for each degree of freedom) is
\be
E=2\sum_{n_1,n_2,n_3}\frac{\hbar \omega_{n_i}}{2}=\hbar \sum_{n_1,n_2,n_3}\sqrt{\frac{n_1^2+n_2^2+n_3^2}{L^2}}.
\ee
But we consider the sum over modes such that the maximum total momentum is $k_{\rm max}=M_{\rm Pl}$, i.e. that the maximum 
$n=\sqrt{n_1^2+n_2^2+n_3^2}$ is $n_{\rm max}=LM_{\rm Pl}$. The sum over $n_i$'s can be approximated by an integral, since it is dominated 
by the upper range, where it is a very good approximation, so we have (putting $\hbar=1$)
\be
E\simeq \frac{1}{L}\int_0^{LM_{\rm Pl}} d^3n n=\frac{\Omega_2}{L}\frac{(LM_{\rm Pl})^4}{4}=\pi L^3M_{\rm Pl}^4.
\ee
The observable part of this should be the energy in terms of $\tilde L$ minus the energy in terms of $L$ (in the absence of modifications to the 
geometry of space or field configuration), and the energy density will be obtained by dividing with the volume of the cube, $L^3$, giving
\be
\rho_\Lambda=\frac{\Delta E}{L^3}=\frac{\pi M_{\rm Pl}^4}{L^3}\left[\left(L+\frac{\a}{M_{\rm Pl}^2L}\right)^3-L^3\right]=\frac{3\a\pi M_{\rm Pl}^2}{L^2}.
\ee

\section{Holographic dark energy cosmology}

In the context of cosmology, we are interested in an expanding sphere of radius $L$, so to compare with the above we consider a sphere of the same 
volume as the cube, thus replacing $L\rightarrow L(4\pi/3)^{1/3}$ and renaming $\rho_\Lambda$ by $\rho_{\rm de}$ (dark energy), so that
\be
\rho_\Lambda\rightarrow \rho_{\rm de}
=\frac{3\a\pi}{(4\pi/3)^{2/3}}\frac{M_{\rm Pl}^2}{L^2}\simeq 3\a\cdot(1.209)\frac{M_{\rm Pl}^2}{L^2}\equiv 3c^2\frac{M_{\rm Pl}^2}{L^2}\;,
\label{rhoL}
\ee
where we have defined a constant $c$ that naturally takes a value close to 1. I will asume $c\simeq 1$ in the following, and drop it.
In a real  (as opposed to this effective one) 
quantum gravitational approach, it is likely that it should be fixed to one, as this is the value that gives the cosmology most consistent 
with observations, as noted in \cite{Li:2004rb}. It remains to decide who is the radius $L$ in the cosmological context. 
As explained in \cite{Li:2004rb}, the choice that works is the future event horizon (the metric is $ds^2=-dt^2+a^2dx^2$)
\be
R_h=a(x(\infty)-x(t))=a\int_t^\infty\frac{dt}{a}=a\int_a^\infty\frac{da}{Ha^2}\;,
\ee
the boundary of a volume a fixed observer may come to observe. It is the one that gives the right cosmology, but also in the context of the calculation 
presented here, it is a natural one to assume for the virtual fluctuations that make up the vacuum energy, as they will give self-consistently the 
future evolution of the Universe.
Another obvious choice, the particle horizon, 
\be
R_H=a\int_0^a \frac{da}{Ha^2}\;,
\ee
does not give the right cosmology.

We should also come back to the original observation of the maximum energy density that can be in a system of size $L$, so that it doesn't collapse
to form a black hole. Given a sphere of radius $L$, the energy $E$ inside it would be bounded by 
\be
E=\rho\frac{4\pi L^3}{3}\leq \frac{L}{2G_N}\;,
\ee
resulting in
\be
\rho\leq \frac{3 }{8\pi G_N L^2}=\frac{3M_{\rm Pl}^2}{L^2}.
\ee
In \cite{Hsu:2004ri,Li:2004rb}, this was taken to mean that the cosmological constant $\rho_\Lambda$ can be only a factor $c^2$ times the maximum 
value, $\rho_\Lambda=3c^2M_{\rm Pl}^2/L^2$, meaning that the vacuum energy reaches its maximum allowable value (up to a factor $c^2$, later taken 
to be 1), but with an $L=R_h$. I think however that given the nature of the bound, it is more logical to consider the {\em total } energy density in the
sphere of size $L$ to be bounded as above by the energy that gives a Schwarzschild radius and collapses into a black hole, 
$\rho=\rho_m+\rho_{\rm de}$, and to be given by {\em exactly} the maximum value at the bound, i.e. 
\be
\rho_{\rm total}=\rho_m+\rho_{\rm de}=\frac{3M_{\rm Pl}^2}{L^2}.
\ee
Here $\rho_m$ is the matter density at the current epoch, but in general contains all other forms of energy in the Universe.
But given the Friedmann equation (\ref{friedmann}), we see that we {\em must} have $L=H^{-1}$, i.e. the length $L$ cannot be anything other than 
the (instantaneous) horizon size. Reversing the logic, for the total energy to be at the limit to create a black hole, the size to be considered must be 
the instantaneous horizon size, since to collapse or not is a decision that is taken at a given moment, and then we derive the Friedmann equation. 

The two {\em independent} equations (\ref{friedmann}) and (\ref{rhoL}) with $L=R_h$ determine the cosmology {\em self-consistently}, if we moreover 
consider an appropriate scaling for $\rho_m$. At the current, matter+$\Lambda$-dominated epoch, we have 
$\rho_m=\rho_0 a^{-3}$. The cosmology is determined
self-consistently, since in (\ref{rhoL}) we have $R_h$, which depends on the future evolution, including (as we will see, rather dominantly) 
on the existence of a cosmological constant (\ref{rhoL}). But then the issue is where does the initial value, determining the current $\rho_\Lambda$, 
come from? As we will see, it comes from the inflationary time, so we must consider the full evolution of the Universe to determine it. 
The exact evolution during the current, matter+$\Lambda$, dominated epoch was found in \cite{Li:2004rb}, but we will not consider it here, since 
we are interested in approximative methods that can be used for the whole evolution of the Universe. 

In order to estimate $R_h$ during the various epochs, we assume that we can approximate the evolution by clean transitions between the various 
regimes: inflation, radiation dominated, matter dominated, $\Lambda$ dominated. 

During the $\Lambda$ dominated epoch starting about now, $a\simeq a_0 e^{eH_0(t-t_0)}$, so 
\be
R_h^{\Lambda D}=a\int_t^\infty\frac{dt}{a}\simeq a_0e^{eH_0(t-t_0)}\int_t^\infty\frac{dt}{a_0 e^{H_0(t-t_0)}}\simeq \frac{1}{H_0}.
\ee
In this epoch, 
\be
\rho_{\rm de}\simeq {\rm const}(t)
\ee
is really like a cosmological constant. 

During the matter dominated (MD) epoch, with the time of the (sharp) transition to $\Lambda$ dominated $t_0$, we have $a\simeq a_0(t/t_0)^{2/3}$, 
whereas as before, during the $\Lambda D$ epoch, $a\simeq a_0 e^{H_0(t-t_0)}$, so 
\bea
R_h^{MD}&=& a\int_t^\infty \simeq a\left[\int_t^{t_0}\frac{dt}{a_0(t/t_0)^{2/3}}+\int_{t_0}^\infty\frac{dt}{a_0e^{H_0(t-t_0)}}\right]\cr
&\simeq &\left(\frac{t}{t_0}\right)^{2/3}\left[3t_0\left(1-(t/t_0)^{1/3}\right)+\frac{1}{H_0}\right]\cr
&\simeq & \left(\frac{t}{t_0}\right)^{2/3}\left[3t_0+\frac{1}{H_0}\right].
\eea
Note first that $t_0\sim 1/H_0$, so the two terms are equally important, and second that, since during matter domination $H=2/(3t)$, 
we have $R_h\gg 3t/2=H^{-1}$. In this epoch,
\be
\rho_{\rm de}\propto \frac{1}{R_h^2}\propto \frac{1}{t^{4/3}}\propto \frac{1}{a^2}
\ee
is not a constant, but decreases slower than matter, and moreover
\be
\left(\frac{\rho_m}{\rho_{\rm de}}\right)_{MD}=R_h^2H^2=\left(\frac{2R_h}{3t}\right)^2\simeq \left[2\left(\frac{t}{t_0}\right)^{2/3}\left(\frac{t_0}{t}+\frac{1}{H_0t}
\right)\right]^2\gg 1\;,
\ee
but is $\propto t^{-4/3}$, so the ratio is decreasing.

During the radiation dominated (RD) epoch, with the time of the (sharp) transition to matter dominated $t_1$, we have $a\simeq a_1(t/t_1)^{1/2}$, 
so we now obtain
\bea
R_h^{RD}&\simeq&a\left[\int_t^{t_1}\frac{dt}{a_1(t/t_1)^{1/2}}+\int_{t_1}^{t_0}\frac{dt}{a_0(t/t_0)^{2/3}}+\int_{t_0}^\infty\frac{dt}{a_0e^{H_0(t-t_0)}}\right]\cr
&\simeq & \left(\frac{t}{t_1}\right)^{1/2}\left\{2t_1\left(1-(t/t_1)^{1/2}\right)+\left(\frac{t_1}{t_0}\right)^{2/3}\left[3t_0\left(1-(t_1/t_0)^{1/3}\right)+\frac{1}{H_0}\right]
\right\}\cr
&\simeq & \left(\frac{t}{t_1}\right)^{1/2}\left[2t_1+\left(\frac{t_1}{t_0}\right)^{2/3}\left(3t_0+\frac{1}{H_0}\right)\right].
\eea
We first note that again, the comparable terms with $3t_0$ and $1/H_0$ dominate, and that $R_h\gg 2t=H^{-1}$ (during the radiation dominated epoch
$H=1/(2t)$). Also during this epoch
\be
\rho_{\rm de}\propto \frac{1}{R_h^2}\propto \frac{1}{t}\propto\frac{1}{a^2}
\ee
is again not a constant, but decreases slower than radiation and slower than matter, since
\be
\left(\frac{\rho_{\rm rad}}{\rho_{\rm de}}\right)_{RD}=R_h^2H^2=\left(\frac{R_h}{2t}\right)^2\gg 1.
\ee

Finally, we consider the inflationary epoch, during which there is an exponential expansion due to an approximate cosmological constant (approximately
constant potential for the inflaton) with Hubble scale
\be
H_i=\sqrt{\frac{\rho_{\Lambda_i}}{3M_{\rm Pl}^2}}.
\ee
Then similarly, with the time $t_e$ when inflation ends and radiation domination begins (we assume the transition is very quick, as for the other epochs)
\bea
R_h^{infl}&=&a\left[\int_t^{t_e}\frac{dt}{a}+\int_{t_e}^{t_1}\frac{dt}{a}+\int_{t_1}^{t_0}\frac{dt}{a}+\int_{t_0}^\infty\frac{dt}{a}\right]\cr
&\simeq & a_e e^{H_i(t-t_e)}\left[\int_t^{t_e}\frac{dt}{a_e e^{H_i(t-t_e)}}+\int_{t_e}^{t_1}\frac{dt}{a_1(t/t_1)^{1/2}}\right.\cr
&&\left.+\int_{t_1}^{t_0}\frac{dt}{a_0(t/t_0)^{2/3}}
+\int_{t_0}^\infty\frac{dt}{a_0e^{H_0(t-t_0)}}\right]\cr
&\simeq & \frac{1}{H_i}+e^{H_i(t-t_e)}\left(\frac{t_e}{t_1}\right)^{1/2}\left[2t_1+\left(\frac{t_1}{t_0}\right)^{2/3}\left(3t_0+\frac{1}{H_0}\right)\right].\label{horizon}
\eea
We have assumed that at least a few e-folds of inflation have already happened, so that the first term is approximately $1/H_i$ (without exponential 
corrections). We now note that the factor $e^{H_i(t-t_e)}$ is very small, so it could in principle dampen the terms it multiplies, so that 
$R_h^{infl}\simeq 1/H_i$. 

To see which of the two is bigger, we estimate the term with $3t_0$ (the term with $1/H_0$ is of the same order, but we are only interested in 
orders of magnitude), which is $=3t_e^{1/2}t_1^{1/6}t_0^{1/3}$. Even if the $1/H_i$ term dominates at the beginning, at the end $R_h^{infl}$ starts 
to increase dramatically due to the $e^{H_i(t-t_e)}$ becoming close to 1 in the second term, so that {\em at the end of inflation and beginning of 
radiation domination}, 
\be
\rho_{\Lambda, e}=\rho_{\rm de}=\frac{3M_{\rm Pl}^2}{R_h^2}\sim \frac{M_{\rm Pl}^4}{3(M_{\rm Pl}t_e)(M_{\rm Pl}t_1)^{1/3}(M_{\rm Pl}t_0)^{2/3}}.
\ee
With $t_0$ close to the current time, $t_0\sim 10^{10}yrs\sim 4\times 10^{18}s$, $t_1\sim 6 \times 10^{10}s$, and 
\be
t_e\sim \left(\frac{T_i}{10^{14}GeV}\right)^{-2} \times 10^{-34}s\;,
\ee
with $T_i$ the inflation temperature, we obtain
\be
\frac{\rho_{\Lambda,e}}{M_{\rm Pl}^4}\sim 10^{-68}\left(\frac{T_i}{10^{14}GeV}\right)^{2}\left(\frac{10^{10}yrs}{t_0}\right)^{2/3}
\sim (e^{-78})^2\left(\frac{T_i}{10^{14}GeV}\right)^{2}\left(\frac{10^{10}yrs}{t_0}\right)^{2/3}.
\ee
Considering also that
\be
\frac{1}{H_i}=M_{\rm Pl}^{-1}\left(\frac{M_{\rm Pl}}{T_i}\right)^2\sim e^{18}M_{\rm Pl}^{-1}\left(\frac{T_i}{10^{14}GeV}\right)^{-2}\;,\
\ee
we can write
\be
R_h^{infl}\sim e^{18}\left(\frac{T_i}{10^{14}GeV}\right)^{-2}+e^{-H_i(t-t_e)} e^{78}\left(\frac{T_i}{10^{14}GeV}\right)^{-1}\left(\frac{t_0}{10^{10}yrs}\right)^{1/3}\;,
\ee
so we need at least
\be
N_{e,min}=60+\ln\left[\frac{T_i}{10^{14}GeV}\left(\frac{t_0}{10^{10}yrs}\right)^{1/3}\right]\label{nefolds}
\ee
e-folds of inflation for the second term to {\em start} inflation smaller than the first, and only afterwards grow to be larger. 
Note that we have explicitly allowed $t_0$ to be different than the current time of $\sim 10^{10}yrs$, 
since $t_0\sim 1/H_0$ is related to the cosmological constant scale, which we will want to fix later. 

We see then the self-consistency of the evolution: the magnitude of the second term is determined by $1/H_0\sim t_0$, which 
determines the evolution, and thus the eventual size of $\rho_{\Lambda,0}$ (our current cosmological constant). This initial condition 
determining $\rho_{\Lambda,0}$, and thus the ratio of $\rho_{\Lambda,0}$ to $\rho_{\Lambda,i}$ (the energy density during inflation),  
can be described as giving the time when the second term in $R_h$ starts to dominate over the first, given a fixed number of e-folds of inflation. 

But note that the first term in $R_h$, $1/H_i$, gives exactly the energy density (due to the inflaton) during inflation, 
$\rho_{\Lambda,i}=3H_i^2M_{\rm Pl}^2$, so in fact the same calculation of the cosmological constant also gives the inflationary cosmological constant!
Yet we would need the inflationary cosmological constant to be constant during the {\em whole} inflationary period, including the last $N_{e,min}$ e-folds, when 
the second term in $R_h$ dominates, so $\rho_{\rm de}$ decreases drastically. That means that, for a consistent picture, almost as soon as inflation
starts, the second term needs to start to increase, and that the real inflationary cosmological constant simply comes from an independent potential 
for the inflaton. The two $\rho_{\rm de} $'s would be equal or at least comparable at the initial point, and one can understand them as having 
the same origin, i.e. the initial inflationary cosmological constant would have a large component due to the fluctuations on scale $R_h$, but 
immediately afterwards $\rho_{\rm de}=3M_{\rm Pl}^2/R_h^2$ would start decreasing drastically, and be a subleading component, to be relevant 
again only in the current epoch, while the rest of the energy components would behave as usual. 

Therefore, we are led to a {\em physical principle}, that the second term in $R_h$ needs to start to increase immediately after the beginning of inflation
for consistency of the physical picture, which leads to fixing of the ratio of cosmological constants during inflation and now, 
$\rho_{\Lambda,e}/\rho_{\Lambda,0}$, in terms of the number of e-folds, by imposing $N_e=N_{e,min}$, with $N_{e,min}$ in (\ref{nefolds}).
As we see, that fixes the physical scale $H_0\sim 1/t_0$. Reversely, knowing experimentally $H_0$ and $\rho_{\Lambda,0}$ allows us to 
{\em fix } the number of e-folds of inflation to $N_e=N_{e.min}$ in (\ref{nefolds}). Note that we moreover obtain an absolute {\em maximum} for the
number of e-folds of inflation. Certainly we need $T_i\leq T_{\rm Pl}=10^{19}GeV$, leading to 
\be
N_e\leq 71.5\;,
\ee
though for reasonable models the bound is even tighter.

Note that \cite{Li:2004rb} also mentioned the possibility of fixing the ratio of scales in terms of the number of e-folds of inflation, but there was no 
clear physical picture behind the proposal.

Finally, we note that it was shown in \cite{Li:2004rb} that a the exact solution for the evolution of the coupled system matter plus $\Lambda$ 
leads to an equation of state with
\be
w=-0.903+0.104z\;,
\ee
since during pure $\Lambda$ domination, $\rho_\Lambda$ is truly constant (corresponding to $w=-1$), whereas during matter domination, 
$\rho_\Lambda\propto 1/a^2$, corresponding to $w=-1/3$ (in general, we have $\rho\propto a^{-3(1+w)}$). 
Note that this seems to be in agreement with the 2015 Planck data \cite{Ade:2015rim} (Fig.7).
The model can be improved by 
considering interaction with dark matter, etc. (e.g. as in \cite{Pavon:2005yx,Wang:2005jx}).

\section{Conclusions}

In conclusion, generic quantum gravity effects, described effectively by a modified uncertainty relatin with a minimum length, coupled with 
the physical principle that the only observable cosmological constant is the one due to changes in the geometry of space or field configuration, allow
us to obtain a result close to experimental observations. As in the holographic dark energy model, we consider the scale $L$ in $\rho_{\rm de}$
to be the future event horizon $R_h$, leading to a self-consistent cosmological evolution. The scale of $\rho_{\Lambda,0}$ today relative to 
$\rho_{\Lambda,e}$ during inflation is set by the initial conditions for $R_h$. For a consistent picture of inflation, we are led to a physical principle, that 
the second term in $R_h^{infl}$ starts to dominate  just at the beginning of inflation. This fixes $\rho_{\Lambda,0}/\rho_{\Lambda,e}$ in terms of the 
number of e-folds $N_e$ and the scale of inflation $T_i$.

{\bf Acknowledgements} I would like to thank Rogerio Rosenfeld for useful comments on the manuscript. 
My research is supported in part by CNPq grant 301709/2013-0 and FAPESP grant 2014/18634-9.

\bibliography{cosmoconst}
\bibliographystyle{utphys}

\end{document}